\documentstyle[stwol]{article}
\def\be{\begin{equation}}
\def\ee{\end{equation}}
\def\bea{\begin{eqnarray}}
\def\eea{\end{eqnarray}}

\begin{document}

\title{ONE-LOOP EFFECTIVE ACTION IN OPEN-STRING THEORY}

\author{ C. FABRE}

\address{Centre de Physique Th\'eorique, Ecole Polytechnique, 
         91128 Palaiseau, FRANCE}


\twocolumn[\maketitle\abstracts{
This is a summary of the analysis of the one-loop effective   
action in $Z_2$-orbifold compactifications of type-I theory 
presented in references [1,2]. We show how,  for non-abelian 
group factors, the threshold effects are ultraviolet finite 
though  given entirely by a six-dimensional field theory 
expression. We also discuss implications for the equivalence 
between Type-I and Heterotic four dimensional $N=2$ superstring 
theories.}]

\section{Motivations}
\subsection{Gauge Couplings Unification in open-string theory}

Assuming the Minimal Supersymmetric Standard Model spectrum at 
low energy, and using the renormalization group equations, the 
extrapolation of low-energy values of gauge coupling constants 
predicts unification at the scale 
\mbox{$M_{U} \simeq 2.10^{16} \  GeV$}. In string theory, 
unification with gravity implies an additional tree-level relation 
at the string scale $M_{string}$, between $\alpha _{U}$ - the 
fine structure constant at unification - and $G_{N}$, the Newton 
constant. Since we know the values of $\alpha _{U}$ and $G_{N}$ 
this relation can be turned into a prediction for $M_{string}$.

In heterotic string theory, a single closed-string diagram 
- the sphere -
is responsible for $\alpha _{U}$ and $G_{N}$ at tree-level.
Then, in the previous unification relation one can eliminate 
the dependence both on the dilaton and on the compactification 
volume, and 
one obtain a model independent prediction for 
$M_{string} \simeq 5.10^{17} \ GeV$ which is bigger than 
$M_{U}$ by a factor $20$.\ ~\cite{DIEN}

In open-string theory $G_{N}$ is given at tree-level by a 
closed-string diagram
while $\alpha _{U}$ is given by an open-string one - the disk , 
so that the
string unification scale depends on the compactification volume.  
Though it is less predictive than the heterotic theory, 
it seems easier to realize tree-level unification in open-string 
theory by adjusting the compactification volume. 
The next step is 
to compute the 
string threshold corrections.  

\subsection{Heterotic/Type-I duality}

There has been some very convincing evidence accumulated 
so far
for the equivalence of theories which were believed in 
the past
to describe truly different types of superstrings. 
Type I, Type II and Heterotic theories seem merely to 
provide 
complementary  descriptions of a more complicated
theory of fundamental interactions, and the larger 
framework of 
superstring 
dualities now includes also M-theory and F-theory 
descriptions.

Among the four-dimensional models, the most familiar 
examples of
dual pairs are based on Type II and Heterotic
constructions.
Type I theory remained a wild card in duality conjectures 
until
quite recently Polchinski and Witten presented several
arguments for the equivalence of Type I and Heterotic
theories in ten dimensions \cite{PW}. Although in $D=10$ 
this is 
a strong-weak coupling duality, it turns out that upon 
appropriate
compactification on $K_{3} \times T^{2}$, one obtains 
four dimensional 
$N=2$ supersymmetric dual pairs with the Heterotic gauge 
coupling
mapped to a Type~I gauge coupling in a way that
some weakly coupled regions overlap on both sides.
One can check their equivalence in the quantitative 
comparisons of
 their low-energy effective actions.

\section{One-loop threshold computation}
\subsection{The open-string model}

In reference [1] we have studied the   
one-loop Lagrangian for slowly-varying 
gauge-field strengths in some $Z_2$-orbifold  models first  
constructed by Bianchi and Sagnotti \cite{BS}, and  analyzed 
in detail  
recently  by  Gimon and Polchinski \cite{GP}. 
These models have N=1  supersymmetry in six dimensions, 
and a maximal gauge group $G= U(16)\times {\tilde U(16)}$,
 which can be broken by both discrete and continuous
(antisymmetric) moduli. Upon toroidal compactification to four
dimensions, one finds  N=2 supersymmetries  and extra (adjoint)
Wilson-line moduli.
The type-I theory on $R^6\times T^4/Z^2$
contains  untwisted and twisted closed strings, 
 as well as open strings of three different types:
those with freely moving endpoints (NN), those whose
endpoints are stuck on some 5-branes transverse to the orbifold
(DD), and  those with one endpoint stuck and one
moving freely (DN). Consistency fixes both the number
of 9-branes and the number of 5-branes to be 32. 
It also fixes the action of the orientation reversal
$(\Omega )$ and orbifold-twist $({\cal R})$
on the open-string end-point states.

The massless spectrum of this theory in six dimensions
 has:

\indent  {\it (i)}  the N=1 supergravity, one tensor
 and four  gauge-singlet
hypermultiplets from the untwisted  closed-string
 sector,\hfil\break
\indent {\it (ii)} sixteen gauge-singlet
  hypermultiplets, one from each
fixed-point of the orbifold, in the twisted closed-string
 sector\hfil\break 
\indent {\it (iii)}  $ U(16)$  vector
 multiplets and two  hypermultiplets in the
antisymmetric $ 120$
 representation from the NN sector,\hfil\break
\indent {\it (iv)} identical  content, i.e.
 an extra ${\tilde U(16)}$
gauge group and two  antisymmetric hypermultiplets,
 from the DD sector,
and \hfil\break
\indent  {\it (v)} one hypermulitiplet transforming in the
representation  $(16,16)$ of  the full gauge group
and coming  from the DN sector.\hfil\break
\indent  Notice that each  twisted-sector
hypermultiplet  contains  a  RR 4-form field $ C^{(I)}$ 
localized at the $I$th fixed point of the orbifold, which 
plays
a special role in what follows.
We will  furthermore choose to  work on 
$ R^4 \times T^2 \times T^4/Z_2 $, 
so that we may break the gauge group by Wilson-line moduli 
in four
dimensions.

\subsection{Method and results}

In calculating the effective gauge-field action
we have followed reference  \cite{BP}, where a  similar 
calculation was carried out  for toroidal compactifications 
of the type-I  SO(32) superstring.
In order to determine the threshold effects
at one-loop compute the free energy in the presence of 
a magnetic background ${\cal{B}}$ 
and then extract the coefficient of ${\cal{B}}^{2}$ 
in the weak-field limit.
This model has a T-duality, which interchanges 
NN and DD sectors
and hence also the two U(16) gauge groups. Without losing
generality, we may thus  restrict ourselves to background 
fields
arising from the NN sector.

The one-loop free energy is the sum of contributions 
from the various
sectors :
\[
{\cal F}^{orbifold} = {\cal F}_{closed}+
{\cal F}_{NN}+{\cal F}_{ND}+{\cal F}_{DD} 
\]
Since only Neumann endpoints couple to the background 
field, 
we can ignore 
${\cal F}_{closed}$ and ${\cal F}_{DD}$ which  vanish
by space-time supersymmetry.
In the remaining terms one find contributions of the 
annulus, 
and the M\"obius strip.
Using the Jacobi identity, one finds that all these 
amplitudes expanded to quadratic order in
the magnetic field collapse to 
contributions  of six-dimensional massless states.
This is to be expected: indeed,  only BPS states
can  contribute to  threshold effects, and the only
BPS states of the  open string are the massless
(before Higgsing)  modes of the
 six-dimensional model. 
A similar result is well known  in 
 the heterotic string \cite{greg}
where, however, the spectrum of
  BPS states includes infinite string excitations with no
simple field-theoretic description.

The final result for the full free energy, including 
classical and
one-loop contributions is a sum over annulus and M\"obius 
contributions :
\[
\  {\cal F}({\cal B}) /  V^{(4)} =\  
{{\cal B}^2 \over 2 g^2_{(4)}}  
 +\ {{\cal B}^2\over 8\pi^2}\   
 \int_0^\infty {dt\over t}\ \times \ \ \ \ \ \ \ \ \ \ \]
\[ \times \Biggl[ 
 \ \ - \sum_{ij} 
  s_{ij} (q_i+q_j)^2\sum_{a_i+a_j+^*\Gamma_2}
  {1\over 4}  e^{-{\pi t} p^2/2}\ \ \ + \ \     
\]
\[ \sum_i  q_i^2 \Bigl(\sum_{a_i+^*\Gamma_2}
 4  e^{-{\pi t} p^2/2}  - \sum_{2a_i+^*\Gamma_2} 
e^{-{\pi t} p^2/2}\Bigr)
\Biggr]
 +o({\cal B}^4)
\]
\\
\noindent 
where the SU(16) generators are 
 normalized to $tr_{16} Q^2 = {1\over 2}$,
$a_i$ the Wilson-line moduli, 
and we recall that the Chan-Patton charges $q_i$ run 
over both the 
$16$ and the ${\overline {16}}$ representations 
separately.  
The orbifold-twist operator acts as a simple sign, 
 $$s_{ij}= -1\ \ \ {\rm  or} \ \ \ +1 \ , $$
according to whether the end-point states
$\vert i>$ and $\vert j>$  belong to  the same or
 to conjugates representations.

As a check let us extract the leading infrared 
divergence of
the coupling-constant renormalization
in the limit of vanishing Wilson lines. Cutting
off $t < 1/\mu^2$ one finds after some straightforward 
algebra
$$
{4\pi^2\over g_{(4)}^2} \Biggl\vert_{1-loop} =
{4\pi^2\over g_{(4)}^2} \Biggr\vert_{tree} - 6\ 
 \log\mu +\  {\rm IR\ finite}
\ 
$$
in agreement with the correct $\beta$-function 
coefficient 
of the
N=2 theory in four dimensions,
$$
C_{adj} - 2 C_{120} - 16 C_{fund} = -6 \ . 
$$

\subsection{What about ultraviolet behaviour ?}

Considered as an expression of field theory, the 
threshold effects
would be quadratically divergent in the ultraviolet 
limit. 
How then is the string result finite ?

In heterotic string theory ultraviolet finiteness at 
one loop follows 
from the restriction of the integration  over all  
world-sheet  tori
to a single  fundamental domain -thanks to modular 
invariance. This 
presupposes conformal invariance, or equivalently 
the absence of 
classical tadpoles.
In open-string theory we do not have modular invariance 
any more but
we learn from references [12]
that the ultraviolet divergences come from tadpoles 
of the massless 
closed string states sandwitched with an on-shell 
propagator.

Compared with the toroidal case, the orbifold model 
has only
one additional piece of potential ultraviolet 
divergence.
It comes from the twisted RR 4-forms, which couple 
to the 
background field through
the generalized Green-Schwarz interaction ~\cite{DOUG}
$$ (2\pi)^{-5/2} \sum_I \int d^6x \ {1\over 4!\ 2}
\epsilon^{\mu\nu\rho\sigma\kappa\tau} 
C^{(I)}_{\mu\nu\rho\sigma}
tr F_{\kappa\tau} \ $$ for canonically-normalized 
4-forms. 
The coupling gives mass to the $U(1)$ (abelian)  
gauge field,
rendering a  background inconsistent. 
For non-abelian background fields, on the other hand, 
the structure of ultraviolet divergences is identical 
to that  
of the toroidal model. Furthermore since in the 
toroidal theory
the gauge coupling is not renormalized, we may 
conclude 
that in the orbifold the renormalization is 
ultraviolet finite.

The first physical divergence appears at order 
$o({\cal B}^4)$
and comes form tadpoles of order $o({\cal B}^2)$ 
of the graviton 
and the dilaton.
It can be used to derive the relation between gauge 
and gravitational
couplings : taking into account the halving of the 
volume, 
we may see that, with SO(32) normalizations for 
the generators of U(16), the relation between gauge 
and gravitational
couplings remains the same in the orbifold model as 
in the toroidal one.

One may be puzzled by the fact that our final result 
for the free 
energy is formally identical to  that of  Kaluza-Klein 
theory  
compactified from six to four dimensions, and yet is 
ultraviolet finite.
If we were to  impose a uniform  ultraviolet
 momentum cutoff,  $ t> 1/ \Lambda^{2}$, the result 
would
indeed be quadratically divergent. The cutoff dictated 
by string 
theory is,  however,  much smarter ! It is uniform in  
transverse time $l$.
The relation between $l$ and the proper time in the 
direct channel
is  different for each  surface 
$$
 l = \cases{& $1/t$ \ \ annulus ;\cr
&$1/4t$ \ \ M\"obius ;\cr
& $1/4t$ \ \ Klein bottle .\cr} 
$$
\noindent 
so that if we cutoff the  annulus at $ t= 1/ \Lambda^{2}$,
we must cutoff   the M\"obius strip at  $t= 1/ 4\Lambda^2$.
The ultraviolet divergence vanishes as we sum over 
the annulus and 
the M\"obius contributions in the free energy.\\
%\\[\baselineskip]

Our  calculation illustrates in a very simple context 
the way 
in which open-string theory produces  finite answers: 
in the case at hand it is simply field theory but with 
a very smart  
cutoff on the momenta.

\section{Heterotic - Type I duality} 

These results have non-trivial implications, 
both in the context of heterotic-type-I duality 
\cite{ABFPT},
and for the study of moduli spaces of D-branes \cite{dyn}.
In this latter context in particular they  imply  
that  the metric in the moduli space of N=2 configurations 
of D-branes
is given entirely by simple and massless closed-string 
exchange.

We now give one specific model which has simultaneous 
Type I, 
Heterotic and type II descriptions. On the type I side 
it originates 
from a six-dimensional model with one tensor multiplet 
and a completely 
broken gauge group. This model can be obtained from the 
class of 
orientifold constructions that is presented in 
section (2.1).
Upon toroidal compactification to $D=4$ one finds the 
3 universal
vector multiplets $S$, $S'$ and $U$.
A Heterotic-Type II dual pair with the same massless 
spectrum has
been considered before in refs.[9] - the so called 
$S-T-U$ model.

Thanks to $N=2$ supersymmetry, the low-energy effective 
actions 
are entirely given by the respective prepotentials, 
which are well 
known for the Heterotic-Type II dual pair.
For the Type I construction the tree-level prepotential is
\[
F^{(0)}=SS'U-{\textstyle\frac{1}{2}\sum_i}
(SA_i^2 + S'A^{'2}_i)
\]
where $\vec{A}$ and $\vec{A'}$ refer to the open-string 
gauge 
multiplets from NN and DD sectors respectively,
$S$ and $S'$ are the couplings to this two kind of 
gauge field and 
$U$ is the usual complex modulus which determines 
the complex 
structure of the torus.
The complex scalars $S$, $S'$ and $U$ belong to vector 
multiplets, and
parametrize a $[SU(1,1)]^3$ manifold in the abscence of 
open-string
vector multiplets. 
We can deduce from Type I-Heterotic duality in $D=10$ 
the corresponding duality in four dimensions:
\[
S_I=S_H\qquad\qquad S'_I=T_H\quad\qquad U_I=U_H\, 
\]
Here, $S_H$ is the heterotic dilaton and $T_H$ 
is the usual K\"ahler structure modulus of the torus.	

The type-I theory exhibits two continuous Peccei-Quinn 
symmetries 
associated to $S$ and $S'$ which dictate the following 
form
of Type I prepotential:
\[
F(S,S',U,A)= F^{(0)}+f_I(U,A)+ 
{{\mbox{non-perturbative}}\atop{\mbox{corrections}}}
\]
where $f_I(U,A)$ is the one-loop correction. 
Type-I non-perturbative terms are suppressed in 
the region we consider.
 
In order to determine perturbative corrections 
to the prepotential, one could in principle follow the 
method 
applied on the Heterotic side, by extracting the 
one-loop 
K\"ahler potential $K^{(1)}$
from the universal part of threshold corrections
to gauge couplings $\Delta$ \cite{ANGANA}.
But unlike the Heterotic case, in Type I theory there 
are also one-loop 
corrections to the Planck mass $\delta$ 
so that the one-loop K\"ahler potential $K^{(1)}$ is 
given by :
\[
\partial_U\partial_{\bar{U}}\Delta
+ \frac{1}{2S'_{2}} \sqrt{G} \;
\partial_U\partial_{\bar{U}}\delta\ 
=-\frac{b}{(U-\bar{U})^2}
+K^{(1)}_{U\bar{U}}\ 
\]
where $b$ is the one-loop beta function coefficient.
Threshold computation has been presented in section (2).
The one-loop correction to Planck mass can be calculated 
in
a similar way~\cite{ABFPT}, and is related to an 
index of 
the Ramond open-string sector.

As expected from the superstring triality conjecture, 
we have 
shown~\cite{ABFPT} that all three prepotentials 
agree in
the appropriate limits : 
\[ Im(S) > Im(S') \rightarrow \infty \]
in the open-string case and the corresponding region : 
\[ Im(S) > Im(T) \rightarrow \infty \]
 in the Heterotic one.

\section*{Acknowledgments}

I thank C. Bachas for a careful reading of the 
manuscript.
This research has been partially supported by 
E.E.C. grant CHRX-CT93-0340, 
and the French-Polish cooperation project 6460.

\end{document}